\documentclass[useAMS,usenatbib,fleqn,twocolumn]{aastex63}
\usepackage{amsmath}
\usepackage{multirow}
\usepackage{graphicx}
\usepackage{color,soul}
\usepackage{epsfig}
\usepackage{amssymb,amsmath}
\usepackage{aas_macros}
\usepackage{hyperref} 
\hypersetup{colorlinks,citecolor=blue,urlcolor=magenta,bookmarks=false,hypertexnames=true} 
\usepackage{verbatim}	
\usepackage{xcolor}
\newcommand{\kms}{km\,s$^{-1}$}
\newcommand{\msun}{$\rm M_{\odot}$}
\newcommand{\uJy}{$\rm\mu Jy\,beam^{-1}$}
\newcommand{\mJy}{$\rm mJy\,beam^{-1}$}
\newcommand{\whz}{W\,Hz$^{-1}$}
\newcommand{\ergs}{$\rm erg\,sec^{-1}$}
\newcommand\fhr{\mbox{$\hspace{0.1cm} \!\!^{\mathrm h}$}}%
\newcommand\fmin{\mbox{$\hspace{0.1cm}\!\!^{\mathrm m}$}}%
\usepackage{soul}
\usepackage{mathrsfs,amsmath}
\begin{document}
\shorttitle{S-shaped GRG (SGRG): J0644$+$1043}
\title{Discovery of 100\,kpc narrow curved twin jet in S-shaped giant radio galaxy: J0644$+$1043}

\shortauthors{Sethi Sagar et al.}
\correspondingauthor{Sagar Sethi}
\email{sagar.sethi@doctoral.uj.edu.pl}

\author[0000-0001-8561-4228]{Sagar Sethi}
\affiliation{Astronomical Observatory, Jagiellonian University, ul. Orla 171, 30-244 Krakow, Poland}

\author[0000-0002-3097-5605]{Agnieszka Ku\'zmicz}
\affiliation{Astronomical Observatory, Jagiellonian University, ul. Orla 171, 30-244 Krakow, Poland}
\affiliation{Queen Jadwiga Astronomical Observatory in Rzepiennik Biskupi, 33-163 Rzepiennik Strzy\.zewski, Poland}

\author[0000-0002-0870-7778]{Marek Jamrozy}
\affiliation{Astronomical Observatory, Jagiellonian University, ul. Orla 171, 30-244 Krakow, Poland}

\author{Lyuba Slavcheva-Mihova}
\affiliation{Institute of Astronomy and NAO, Bulgarian Academy of Sciences, 72 Tsarigradsko Chaussee Blvd., 1784 Sofia, Bulgaria}

\begin{abstract}
We report the discovery of an S-shaped morphology of the radio galaxy J0644$+$1043 imaged with a 30 \uJy \,sensitive 525\,MHz broadband (band 3 $+$ 4) uGMRT map. Dedicated spectroscopic observations of the host galaxy carried out with the 2-meter Rozhen telescope yielded a redshift of 0.0488, giving a projected linear size of the peculiar radio structure of over 0.7\,Mpc. This giant radio galaxy is powered by a black hole of mass 4.1$^{+9.39}_{-2.87}\times 10^8$ \msun, from which vicinity emanate well-collimated and knotty jets, each $\sim$100\,kpc long. The entire radio structure, presumably due to the effective jet precession, is less than 50\,Myr old, has a power of $\sim$6 $\times 10^{24}$ \whz \,at 1.4\,GHz and the observed morphological characteristics do not strictly conform to the traditional FR\,I or FR\,II categories.
\end{abstract}


\section{Introduction}\label{sec1:intro}
Jets are common in a wide range of extragalactic sources, and their size can vary from pc to Mpc scales in radio-loud active galactic nuclei (AGN). These jets are highly collimated relativistic outflows that emerge from a supermassive black hole (SMBH, mass range $\rm 10^6-10^{10}$ \msun) located at the centers of galaxies \citep{SMBH.Kormendy.Richstone,SMBH.Kormendy.Ho}. Jets of relativistic charged particles are believed to arise due to the strong magnetic field produced by the rotating accretion disk through the Blanford-Znajek \citep{Blandford.Znajek} or Blanford-Payne \citep[][]{Blandford.Payne.1982} mechanisms. Two different types of jet are observed in radio AGNs: one is a strong-flavored jet \citep[e.g., jets in Cygnus\,A:][]{CygA.RP84}, characterized by being one-sided, well-collimated, narrow, highly relativistic, and terminating with hotspot. The other type is weak-flavored jets \citep[e.g., jets in 3C\,31:][]{3C31.LB02a,3C31.LB08a}, which are two-sided, flaring in nature, subrelativistic, and dissipate into diffuse lobes or tails rather than terminating sharply \citep{Bridle.1984.type.of.jet, Jet.Review.Bridle84}. Morphologically, the extended radio structures created by such jetted AGN were further divided into two distinct types of radio galaxies (RGs): Fanaroff Riley type I and II \citep[FR\,I and II;][]{FR-type}. The jets in both FR type RGs differ structurally and relating types of jets to classical FR types is not straightforward. Initially, the morphological classification of RGs and jets was on the basis of the 3CR and/or $\sim$2\,Jy samples \citep[see, e.g.,][and references therein]{Jet.review.Saikia.2022}{}{}. However, our understanding of the FR dichotomy is changing rapidly due to the observational evidence from the new generation radio surveys \citep[e.g.,][]{Best.heckman.2012,Mingo.FRI.FRII.2019,Mingo.2022}.

Modeling of weak-flavored jets in nearby bright and extended FR I RGs has been successfully carried out many times in the past, mainly due to extremely sensitive Very Large Array (VLA) observations (measurement noise $\leq$ sub \mJy), which easily revealed them in all their glory \citep[e.g.,][]{Jet.review.FRI.Laing14}. However, the strong-flavored jets are difficult to model for two reasons. First, they are slender and thus difficult to resolve at the current sensitivity limits of low-angular resolution radio telescopes/surveys. Second, these jets are either one-sided or contain a very faint counter-jet on the kpc scale due to relativistic aberration. Therefore, it is difficult to disentangle the structure of such jets, which may also be embedded in the surrounding diffuse cocoon/lobe emission \citep[][]{Jet.Laing.SKA.review,Jet.review.Saikia.2022}. Detailed studies of jets are crucial to understanding the evolution of RGs, their host galaxies, and their larger environments.

Jets in some RGs show significant deviations from their initial jet direction, leading to the formation of several distorted types of RGs, that is, a mirror-symmetric one called C-shape and an inversion-symmetric one called S- or Z-shape \citep{S.Shape.Florido.1990,WAT.C.shape.Rudnick.Owen.1976.1,Env.C.S.Liu.2019,WAT.review.O'Dea.Baum}. In the case of C-shaped RGs, their dynamical nature has been well recognized to date. This type of distortion is caused either by the translational motion of the galaxy through the intergalactic medium (IGM) or by the orbital motion of an RG host around a nearby companion galaxy \citep{WAT.C.shape.Rudnick.Owen.1976.1}.  Thus, among C-shape RGs we distinguish three different morphologies: wide-angle tailed \citep[WAT, e.g., 3C\,465;][]{WAT.Eilek.1984.What.bends.3C.465}, narrow-angle tailed (NAT) and/or head-tail \citep[HT, e.g., 3C\,129;][]{HT.2002.Lane.W.M.3C129}. The S- and Z-shape of an RG can result from jet precession, which could be explained in two ways: (i) the presence of a binary black hole at the nucleus \citep[e.g.,][]{Begelman.BH.bainary} and/or (ii) the occurrence of the massive tilted accretion disk \citep[e.g.,][]{Lu.tilted.acc}. However, because of the complex conditions that give rise to various structures, the detailed formation of different morphological types of RGs has not yet been investigated.
 
RGs can attain very large sizes, with those larger than 0.7 Mpc being referred to as giant RGs \citep[GRGs; for references see e.g.,][]{Kuzmicz18, Pratik.SAGAN, PratikLoTSS, Dabhade2023}. The morphology of most known GRGs falls into the FR II category, while only a small percentage (from 4.9 to 7.5\%) falls into the FR I category \citep[][]{Saikia.GRS.1999,Kuzmicz18,Pratik.SAGAN,PratikLoTSS}. However, the number of discovered FR I-type GRGs may increase significantly in the future, thanks to planned observations sensitive to extended diffuse radio structures at low frequencies. 

This paper presents new sensitive radio observations of the peculiar structure of J0644$+$1043, revealing an extraordinary S-shaped morphology together with a 100 kpc twin jet. This jet can be considered a ``naked jet'', as no diffuse radio cocoon has been detected around it. This RG also shows mixed FR-class behavior, which we discuss in later sections. The structure of the paper is as follows. We present selected archival data in Section\,\ref{sec2:archival}, the new radio and optical observations and their reduction in Section\,\ref{sec3:Obs}, while the results are presented in Section\,\ref{sec4:result}, followed by discussion and conclusions in Section\,\ref{sec5:discussion} and \ref{sec6:conclusion}, respectively.

Throughout the paper, we adopt the flat $\Lambda$CDM cosmological model based on the Planck Collaboration \citep[$\rm H_0 =67.8$\,\kms\,Mpc$^{-1},\rm \Omega_m$= 0.308;][]{Plank2016}. We use the convention $\rm S_{\nu}\propto \nu^{\alpha}$, where S$_{\nu}$ is the flux density at frequency $\nu$ and $\alpha$ is a spectral index. The flux density scale is that of \citet{Baars.1977}. All positions and maps are given in the J2000.0 coordinate system.

\begin{figure*}[!ht]
\includegraphics[scale=0.4]{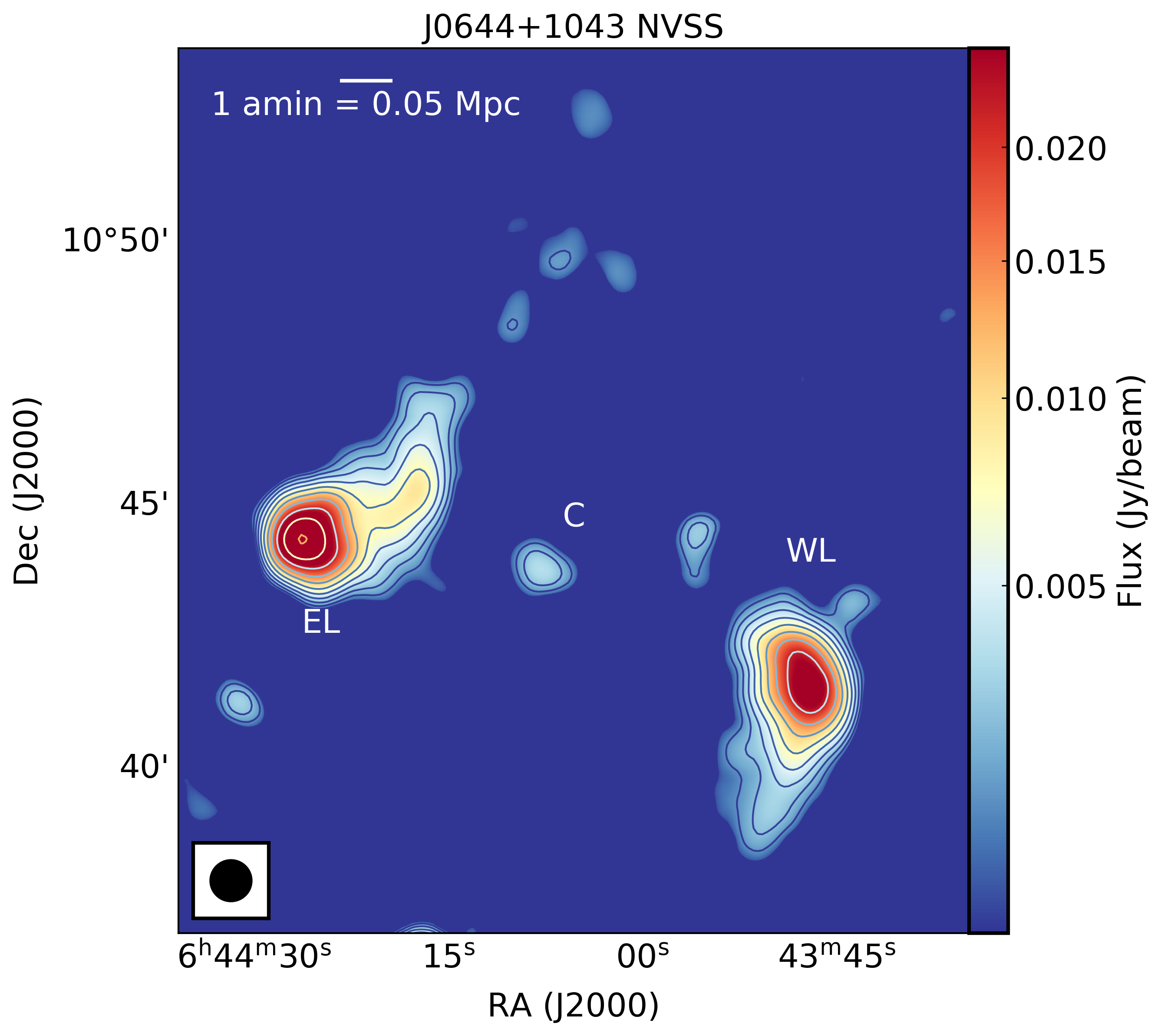}
\includegraphics[scale=0.4]{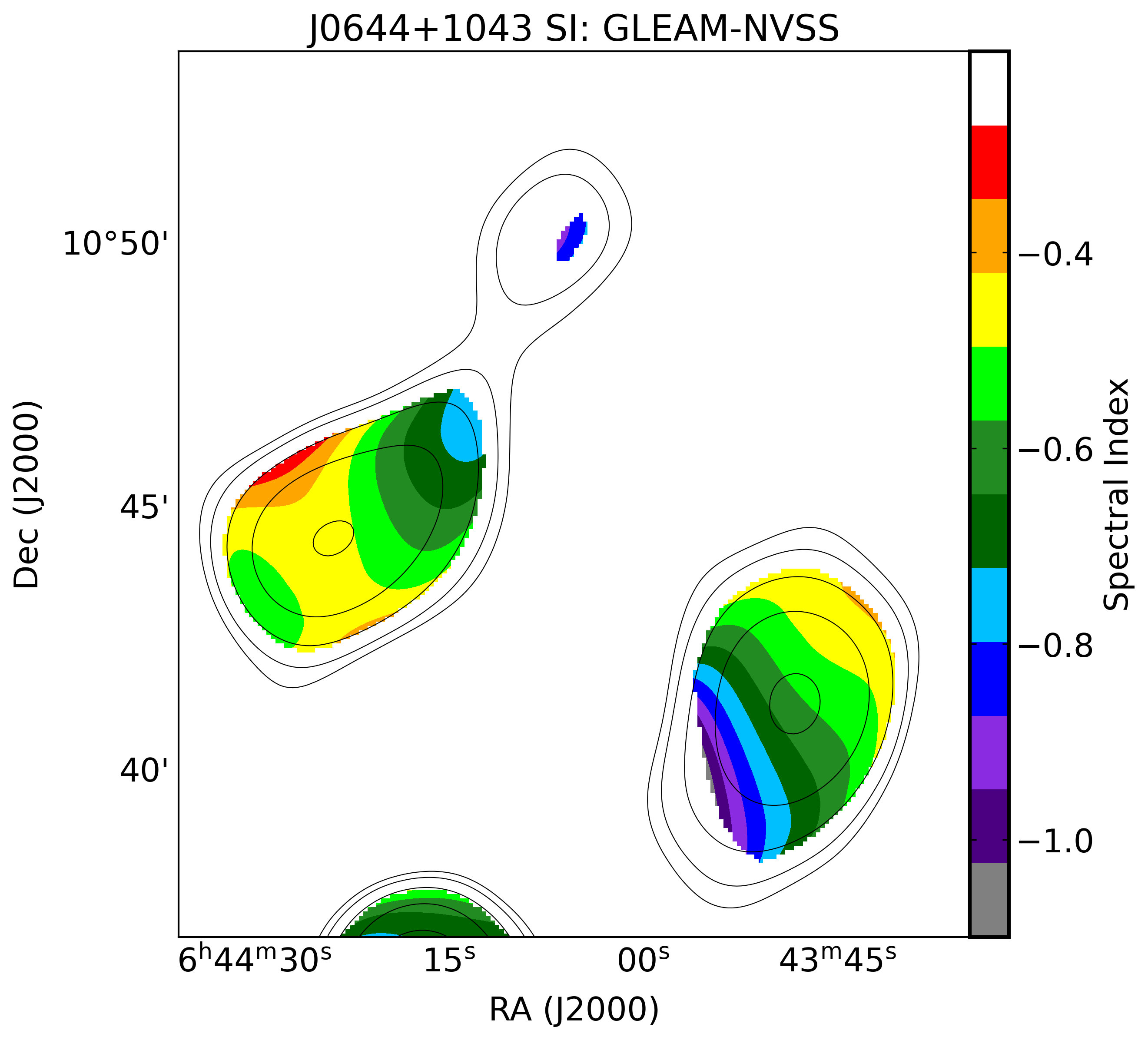}
\caption{\texttt{J0644$+$1043} archive data: The left panel shows the 1400 MHz NVSS radio map plotted above three sigma level ($\rm \sigma$= 0.45 \mJy). The NVSS contours are plotted at $\rm 3\,\sigma \times \sqrt{2} ^n$ (n = 0, 1, 2, 3 ...). A black circle in the lower left corner indicates the beam size. On the NVSS map we have also marked the western (WL) and eastern (EL) lobes and the core (C). The angular to linear length scale is given at the top of figure and subsequent figures in the article. The right panel shows the 200\,MHz GLEAM  contours on the SI map between 200\,MHz GLEAM and the convolved 1400\,MHz NVSS maps. The GLEAM contours are plotted at $\rm 3\,\sigma \times 2^n$ (n = 0, 1, 2, 3 ...).}
\label{fig:nvss_SI}
\end{figure*}

\section{J0644$+$1043 in the literature and archival data} \label{sec2:archival}
J0644+1043 was originally cataloged by \citet[][]{Proctor.2016}, who selected it as an irregular/distorted GRG candidate based on its considerable angular size. Then \citet[][hereafter referred to as D20]{Pratik.SAGAN} identified the host galaxy of J0644$+$1043 with apparent r-band magnitude of 14.56. Using its photometric redshift \citep[z$_{\rm phot}$= 0.24; taken from][]{redshift.s.grg.pratik} and angular size of $\sim$ 12\arcmin, D20 estimated its linear size as 2.8\,Mpc. However, \citet[][]{Bilicki.redshift} reported a significantly lower estimate of photometric redshift, z$\rm _{phot}$=0.03425, resulting in a much smaller linear size for this RG of $\sim$ 0.6 Mpc. This RG is hosted by an elliptical galaxy, WISEA J064408.04$+$104341.5 (RA: 06\fhr44\fmin08\fs04 Dec: 10\arcdeg43\arcmin41\farcs40). It is also listed as a ``zone of avoidance" galaxy by \citet[][galactic latitude b= 3\fdg2]{zone.of.avoid} and is located behind the dark region of the Cone Nebula, where the Gunn r-band galactic extinction is 0.66 magnitude \citep[][]{Galactic.Extinction.S}.

The observational data of J0644$+$1043 are available in many surveys: (i) optical -- Panoramic Survey Telescope and Rapid Response System \citep[PanSTARRS;][]{Pan-STARRS.DR1.2016},  Sloan Digital Sky Survey \citep[SDSS;][]{sdss.York.2000,SDSS.Alam.2015},  Digitized Sky Survey POSS-II (DSS2), (ii) infrared -- Wide-field Infrared Survey Explorer \citep[WISE;][]{wise} and (iii) radio -- Green Bank \citep[87GB, 5\,GHz;][]{87GB.1991}, Very Large Array (VLA) Sky Survey \citep[VLASS, 3\,GHz;][]{VLASS}, NRAO VLA Sky Survey \citep[NVSS, 1.4\,GHz;][]{NVSS}, Rapid Australian Square Kilometer Array Pathfinder (ASKAP) Continuum Survey-low \citep[RACS-low, 0.9\,GHz;][]{RACS,RACS2.Hale} and GaLactic and Extragalactic All-sky Murchison Widefield Array (MWA) survey \citep[GLEAM, 72-230\,MHz;][]{gleam}.

The NVSS map shows this GRG as an FR\,II-type object with an east-west orientation and a hint of extended plume-like lobes that are nearly perpendicular to the jet axis (see the NVSS map of J0644$+$1043, which is shown in the left panel of Figure\,\ref{fig:nvss_SI}). However, it is not clear from the map whether the two lobes and the faint compact object between them form a single object. This is because no bridge connects the two lobes, whose morphology is clearly different from the classical lobes of RGs. Moreover, due to the low angular resolution of NVSS, the two lobes more closely resemble two separate NAT or HT sources.

The integrated flux densities of different source components, presented in Table\,\ref{tab2:flux_comp}, were obtained using \texttt{CASA}\footnote{CASA: Common Astronomy Software Application} \citep{casa} by manually selecting radio emission regions from the GLEAM, RACS-low, and NVSS maps. We used four broadband GLEAM maps for our analysis. The lobes are well detected in all four broad bands of the GLEAM maps, and no confusing point sources overlap with the lobes of this RG in the GLEAM maps. The flux density errors of GLEAM, RACS-low, and NVSS were determined using the following equation (e.g., \citealt[]{Mhl21}{}):

\begin{equation}
   \rm \Delta S_{\nu}  = \sqrt{(S_{\nu} \times 0.1)^2 + \left(rms \times \sqrt{\cfrac{area}{beam}}\right)^2}
\end{equation}

We created a spectral index (SI) map using the GLEAM (170$-$230\,MHz centered at 200\,MHz), and the convolved NVSS (see right panel of Figure\,\ref{fig:nvss_SI}). The SI map shows a steeper to flatter SI from the plume edges to the brightest regions. Furthermore, as expected, the average spectral index is $-0.57 \pm 0.17$. This is not very high value of SI which is due to the dominance of small but bright regions around hotspots containing ``fresh'' energetic plasma with SI $\sim -$ 0.5. This may be evidence of this RG's FR\,II-like nature. As expected, the edges of the plumes containing ``old'' less energetic plasma have SI values close to $-$1. The total radio power at 1400 MHz is 5.92 $\times$ 10$^{24}$ \whz calculated from the NVSS total flux density value, and the kinetic jet power at 150 MHz is 2.01 $\times$ 10$^{42}$ \ergs based on the GLEAM measurement. The jet kinetic power was estimated according to the scheme based on a simulation-based analytical model given by \citet[][]{Analytical.Model.RG.Hardcastle18}. We have extrapolated the 5\,GHz core radio flux using 3\,GHz VLASS flux by considering the SI of $-$0.5 and calculated the upper limit\footnote{The core remained unresolved at 3 GHz. It is more likely to be the case that the core flux we obtained could be contaminated by the inner jets. Thus, the actual core power can not exceed the value we obtained. Therefore, we considered this value as the upper limit for nuclear core radio power and the true innermost core power can be obtained by observation of high-resolution VLA or VLBA observation.} of the core radio power as 5.86 $\times$ 10$^{21}$ \whz. In a later section, we also used this core power to distinguish the type of jet flavor as described by \citet[][]{Bridle.1984.type.of.jet}{}{}.

\newpage
\section{New observations and data reduction}\label{sec3:Obs}
The most likely cause of discrepancy between the photometric redhsifts obtained by \citet[][]{redshift.s.grg.pratik} and \citet[][]{Bilicki.redshift} could be observational uncertainties (such as improper calibration of photometry resulting in galaxy template mismatch) due to its location in the galactic plane. We performed dedicated spectroscopic observations to determine its redhift and radio observations to study large-scale radio morphology. In the following, we provide the details of the observations and data analysis.

\subsection{Optical spectrum}\label{subsec3:optical}
\begin{figure}[h!]
\includegraphics[scale=0.7]{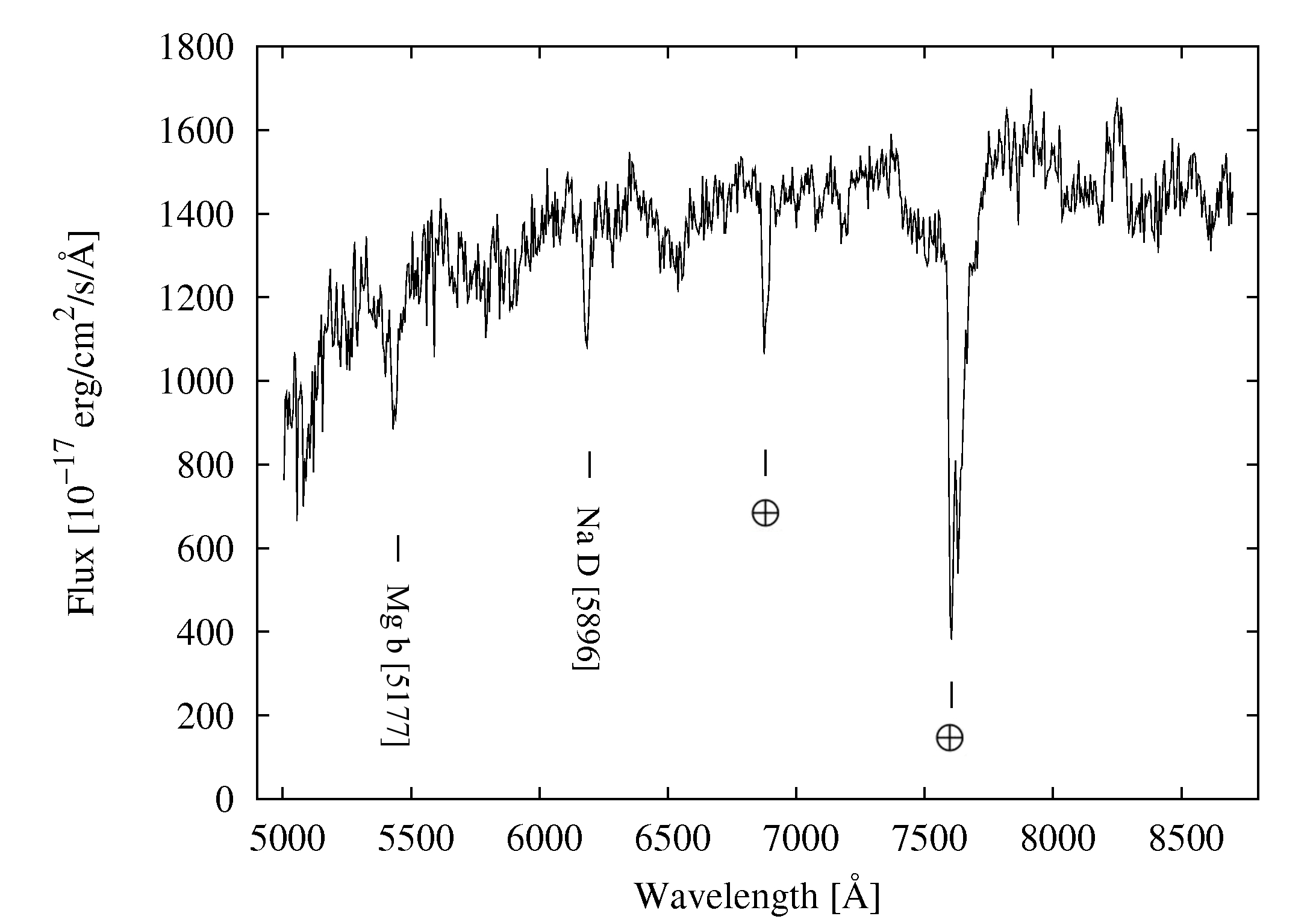}
\caption{Optical spectrum: The spectrum of the host galaxy (RA:\,06\fhr44\fmin08\fs04 Dec:\,10\arcdeg43\arcmin41\farcs40) of the GRG J0644$+$1043 with identified absorption bands. The telluric features are marked with a circled cross mark.}
\label{fig:Opt_Spectrum}
\end{figure}

The spectroscopic observations of the host galaxy of J0644$+$1043 were made on 6/7 February 2022, using the 2-m telescope of the Rozhen National Astronomical Observatory, Bulgaria, equipped with the 2-channel focal reducer FeReRo-2 \citep{Rozen.Jockers.2000}. We used the grism covering the wavelength range from 5000\,{\AA} to 8700\,{\AA} with a dispersion of $\sim$ 5\,{\AA} per pixel (R=6000). Observations were carried out with three exposures of the target galaxy with a total exposure time of 65 minutes. Data reduction was performed using standard procedures with the NOAO Image Reduction and Analysis Facility \citep[\texttt{IRAF}:][]{IRAF86,IRAF93} software, including bias and flat correction, night sky spectrum subtraction, wavelength calibration, and flux calibration. All calibrated spectra of the target galaxy were combined into a final spectrum. The resulting spectrum (presented in Figure\,\ref{fig:Opt_Spectrum}) shows no emission lines. Only absorption features are visible. Two absorption bands were identified: Mg b (5177 {\AA}) and Na\,D (5896 {\AA}). The two prominent absorption features that occur at 6885 {\AA} and 7620 {\AA} are artifacts left after the skyline subtraction procedure corresponding to the telluric O$_2$ molecule.

\subsection{Radio observations and data reduction}\label{subsec3:radio}
The radio observations of J0644+1043 were performed on band 3 and band 4 using the upgraded Giant Meterwave Radio Telescope \citep[\texttt{uGMRT};][]{uGMRT} with the GMRT Wideband Backend (GWB) and simultaneously with the GMRT Software Backend (GSB). The total observing time on the target was 103 and 107 min in bands 3 and 4, respectively. J0521+166 (3C\,138) and J0745+101 were used as phase calibrators, and 3C\,48 or 3C\,147 and 3C\,286 for bands 3 and 4, respectively, were used as flux density calibrators.

The GWB data reduction procedure was carried out by dividing it into smaller frequency sub-bands (6 and 4 sub-bands for bands 3 and 4, respectively), which were calibrated independently for each band using \texttt{SPAM}\footnote{\href{http://www.intema.nl/doku.php?id=huibintemaspampipeline}{\texttt{SPAM}: Source Peeling and Atmospheric Modeling}} \citep[][]{Intema.SPAM.2014, Intema.TGSS.2017}. The analysis starts by automatically flagging bad data, including radio frequency interference (RFI). Flux density and bandpass calibrations are then derived from the respective calibrators, with the calibrator characteristics based on models from \citet[][]{SPAM.Cal.model.Scaife.Heald.2012}. Furthermore, instrumental phase calibrations were also determined using the methods described by \citet[][]{Intema.Ion.Model.2009}. With the use of \texttt{PyBDSF}\footnote{\href{https://pybdsf.readthedocs.io/en/latest/index.html}{\texttt{PyBDSF}: Python Blob Detector and Source Finder} \citep[][]{PyBDSF}}, from the GSB data a reliable sky model was obtained which was then fed into the pipeline for the final step of GWB data reduction. \texttt{SPAM} performs both direction-independent and direction-dependent calibration and imaging. The pipeline underwent several rounds of self-calibration for phase and an amplitude-phase calibration in the final round. Finally, we obtained the calibrated visibilities for each sub-band.

\setlength{\tabcolsep}{2pt}
\begin{table*}[!ht]
    \centering
    \begin{tabular}{ccccccc}
    \hline
    uGMRT & Band & Central &No. of  &  Synthesized & Position &rms \\
    band & width & freq. & sub-bands &beam & angle &($\sigma$) \\
    name &(MHz) &(MHz) &(\arcsec) &(\arcsec $\times$ \arcsec) & (\arcdeg) &(\uJy)\\
    (1) & (2) & (3) & (4) & (5) & (6) & (7)\\
    \toprule
    3 & 200 & 400  &6  & 6.31 $\times$ 5.77 & 16.70 & 60 \\
    4 & 200 & 650 &4  &5.63 $\times$ 4.41 & $-$73.93 & 20 \\
    3$+$4& 400 &525 &10 & 5.76 $\times$ 5.00 &$-$65.59 & 30 \\
    \hline
    \end{tabular}
    \caption{Parameters of the uGMRT observations and final imaging.}
    \label{tab1:wsclean_param}
\end{table*}

\begin{figure*}[!ht]
\includegraphics[scale=0.4]{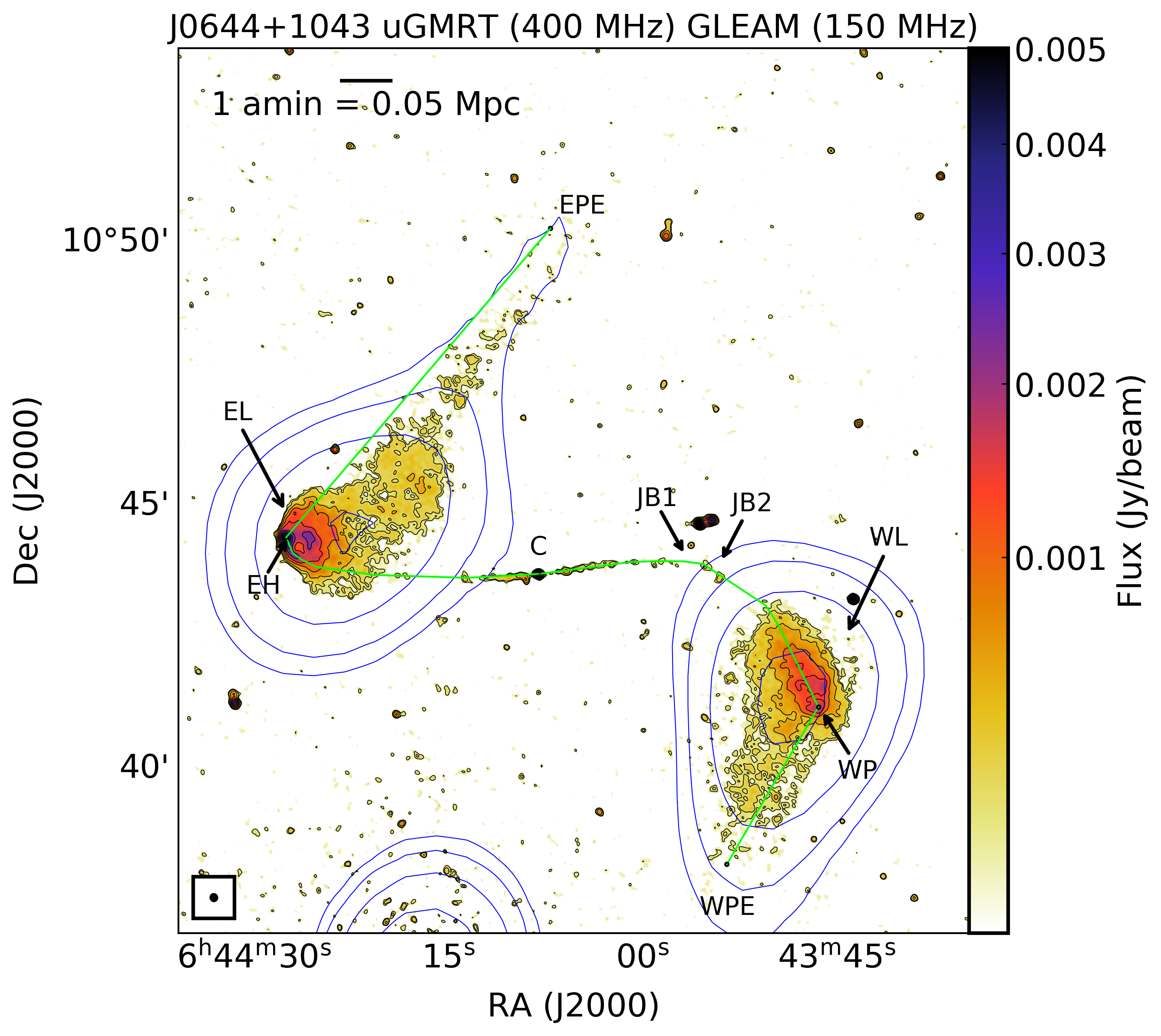}
\includegraphics[scale=0.4]{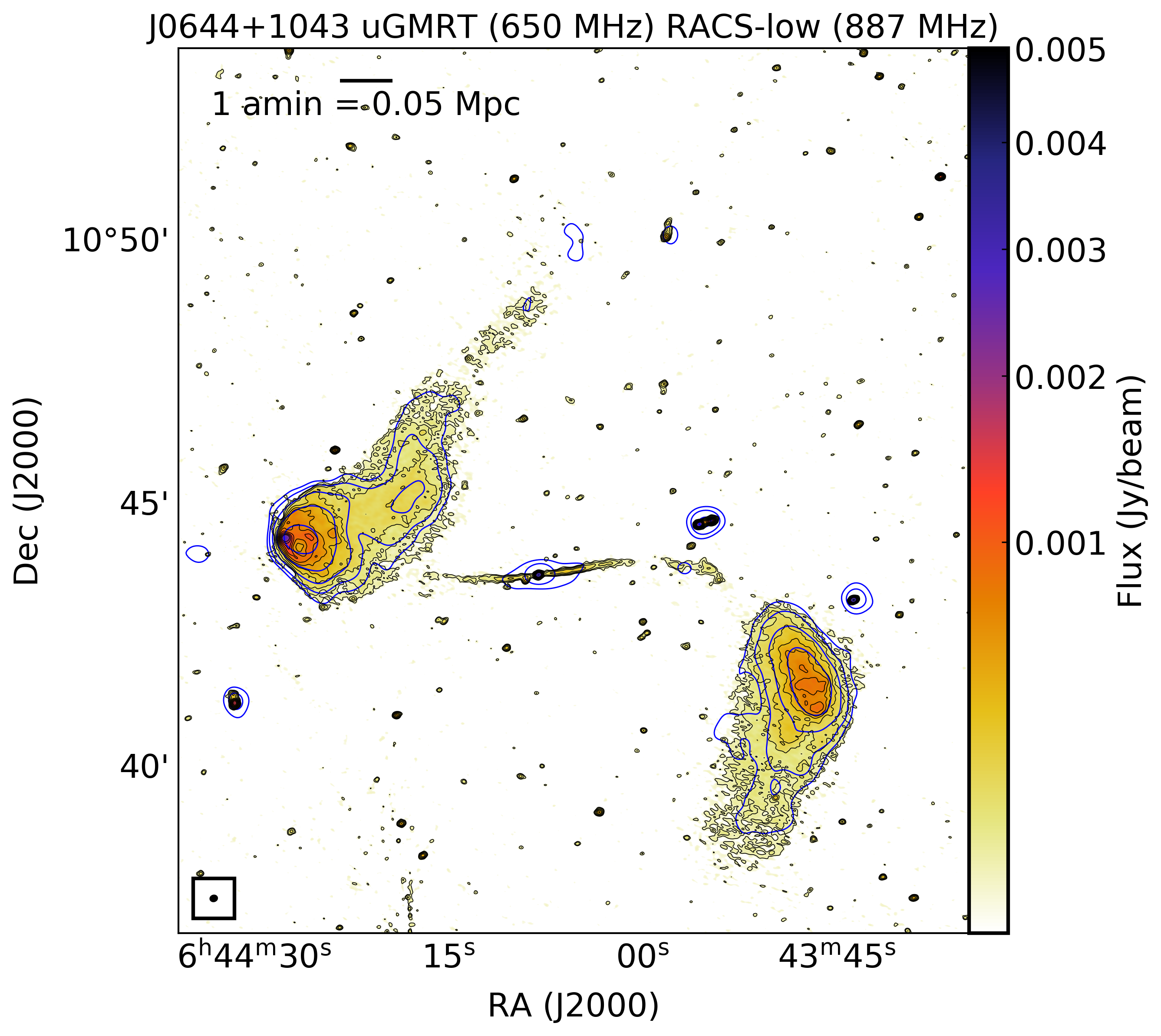}
\caption{\texttt{J0644$+$1043 uGMRT band 3 and band 4 maps}: The left panel shows the 150\,MHz GLEAM map with thin blue contours superimposed on the 400\,MHz uGMRT map with color and black contours. The first contours are at 0.18 and 30 \mJy\, for uGMRT and GLEAM maps, respectively. The GLEAM contours are plotted at $\rm 3\,\sigma \times 2^n$ (n = 0, 1, 2, 3...), while the uGMRT contours are plotted on the square-root scale. The measure components of the RG are described in Section\,\ref{subsec4.2:result} and labeled as follows: EL, C, WL, eastern plume end (EPE), western plume end (WPE), and jet bending (JB1 and JB2). The green line indicates the path through which the angular size is measured. The right panel shows the 887\,MHz RACS-low map with thin blue contours overlayed on the 650\,MHz uGMRT color and black contour map. The first contours are 0.06 and 1.2 \mJy\, for uGMRT and RACS-low map, respectively. In both panels, the uGMRT synthesized beam sizes are marked by a filled circle in the lower left corner of the image. The RACS-low contours are plotted at $\rm 3\,\sigma \times 2^n$ (n = 0, 1, 2, 3 ...) while the uGMRT contours are plotted on the square-root scale.}.
\label{fig:ugmrt_B3_B4_map}
\end{figure*}

The calibrated output visibilities from all sub-bands were then jointly deconvolved in the wideband imager, \texttt{WSClean}\footnote{\href{https://wsclean.readthedocs.io/en/latest/index.html}{WSClean: w-stacking clean}} \citep{WSclean.offringa.2014}, in the form of measurement sets created in the \texttt{CASA}. We used a multi-scale, multi-frequency wideband deconvolution approach \citep[see][]{WSclean.offringa.Smirnov.2017} to image each band.  We chose the appropriate scale (or cell) and image size, respectively, by considering 4-6 pixels in the synthesized beam and the size of the primary beam. To automatically reach the cleaning threshold, the niter parameter is set to a maximum number (here 1,000,000) with the auto-mask and auto-threshold functionality. We used the \citet[][]{Dan.Briggs} weighting to suppress the sidelobes in the point spread function (PSF). A robustness parameter of 0.0 was used, which is a compromise between natural and uniform scales to preserve both large-scale structures and high-resolution details. The wideband multi-frequency deconvolution mode allows channels to be cleaned together, taking into account spectral variation and deconvolving each channel with its own PSF. Primary beam correction was performed using EveryBeam which was implemented in \texttt{WSClean}. Additionally, the residual images were examined to confirm if any emissions were still present and needed further cleaning. Finally, we obtained multi-frequency synthesis (MFS) images at frequencies of 400 MHz for uGMRT band 3 and 650 MHz for uGMRT band 4. The parameters associated with the WSClean imaging and final maps are given in Table\,\ref{tab1:wsclean_param} and Figure\,\ref{fig:ugmrt_B3_B4_map}, respectively.

\subsection{Combined map of uGMRT band 3 and 4}\label{subsec3:525}
\begin{figure*}
    \includegraphics[scale=0.75]{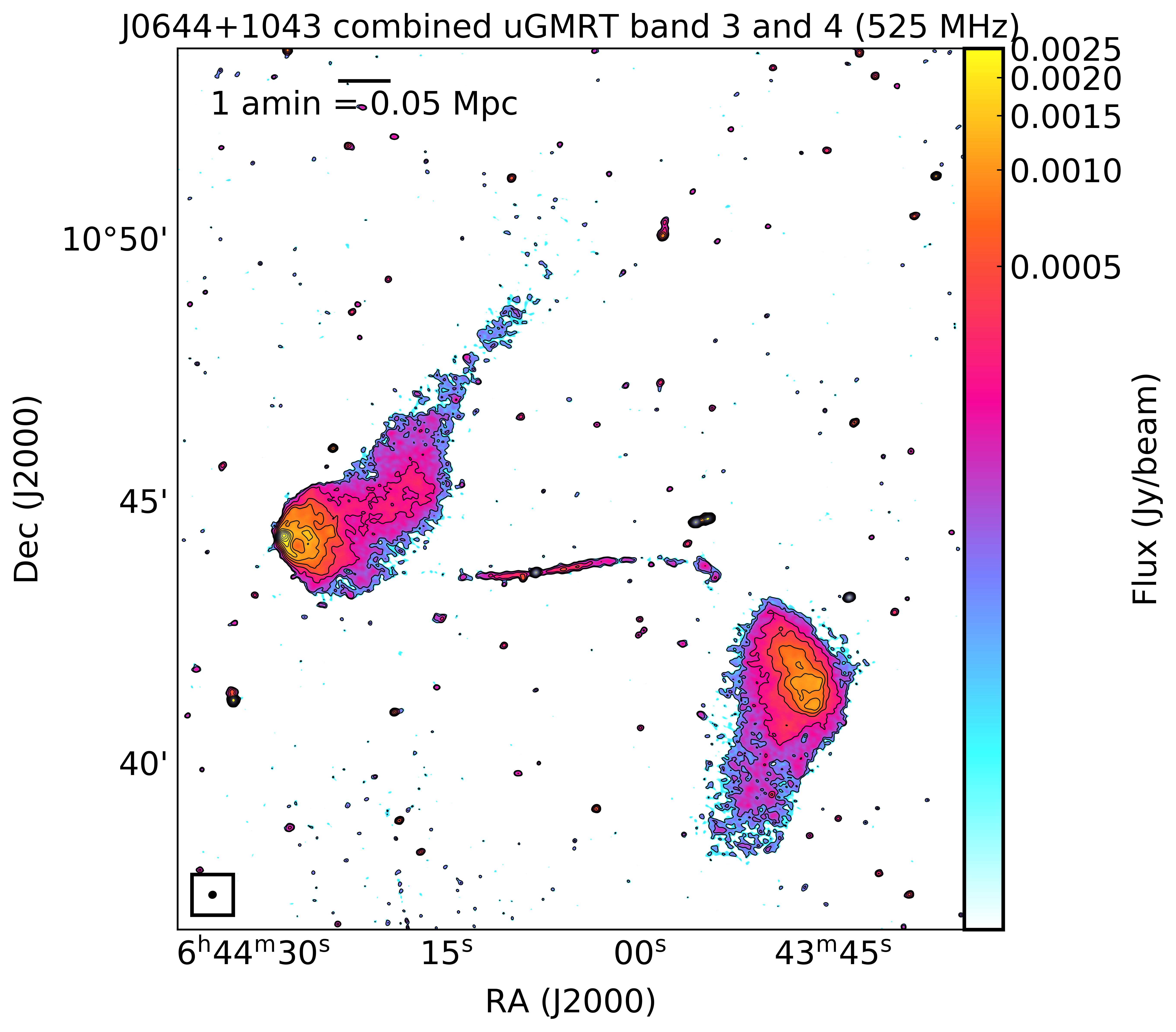}
    \caption{\texttt{J0644$+$1043}: The image shows the combined map of uGMRT band 3 and 4 with the total bandwidth of 400\,MHz centered at 525\,MHz. For the contour plot we have used the square-root scaling to enhance faint radio surface brightness structures present in the source. The first contour is at three sigma level ($\rm \sigma$= 30 \uJy). A filled circle in the lower left corner of the image indicates the size of the synthesized beam. The color scheme created using \texttt{matplotlib} is similar to the RGB1 color scheme of the \texttt{CASA imvew}.}
    \label{fig:uGMRT_map}
\end{figure*}

The new uGMRT maps show a peculiar S-shaped structure. However, the jet structure is better visible with the uGMRT band 4, while the diffuse emission is better imaged with uGMRT band 3. Therefore, it was decided to produce a combined map using observations of band 3 (300 - 500\,MHz) and band 4 (550 - 750\,MHz) to preserve the full structure of the GRG with better uv coverage and sensitivity. The calibrated output visibilities of all sub-bands of band 3 and 4 obtained from \texttt{SPAM} were jointly deconvolved using \texttt{WSClean} as described in the previous section. \texttt{WSclean} performed multi-scale, multi-frequency wideband deconvolution, and we obtained a multi-frequency MFS image at a frequency of 525\,MHz with a bandwidth of 400\,MHz. The spectral variation across the whole band is handled by the wideband multi-frequency deconvolution mode. Due to a total observation time of 6 hours (210 min on source time), the use of two spectral windows results in an image with better sensitivity than in the case of single bands. The resulting source structure is shown in Figure\,\ref{fig:uGMRT_map}, and its parameters are given in Table\,\ref{tab1:wsclean_param}. To the best of our knowledge, this is the first attempt to obtain such a broadband low-frequency map using uGMRT. We will refer to this map as the uGMRT 525 MHz map.

\section{Results}\label{sec4:result}
\subsection{Redshift, velocity dispersion, and black hole mass}\label{subsec4:redshift}
We estimated the redshift $\rm z_{spec}$ value to be 0.0488$\pm$0.0035 by cross-correlating the observed spectrum with the template spectrum of an early-type galaxy. We used the routine described by \citet[][]{Tonry.Davis.1979}. The spectroscopic redshift is five times lower than the photometric redshift of \citet[][]{redshift.s.grg.pratik} mentioned in D20 and slightly higher than ($\rm z_{phot}$=  0.0342) estimated by \citet[][]{Bilicki.redshift}. Using the obtained spectrum of the J0644$+$1043 host galaxy, we determined the stellar velocity dispersion ($\rm \sigma^*$) and the black hole mass ($\rm M_{BH}$). To do this, we corrected the observed spectrum for galactic extinction, A$_V$, using data from the NASA/IPAC Extragalactic Database. The extinction-corrected spectrum was then transformed to the rest frame using the estimated redshift value. We applied the simple stellar population synthesis code STARLIGHT \citep[][]{SSP.2005MNRAS} to model the observed spectrum by fitting a galaxy spectral continuum. As a result, we obtained a stellar velocity dispersion $\sigma^*$= 379 \kms. The quality of the stellar continuum fit is measured by the reduced $\chi^2$ and the {\it adev} parameter, which is the percentage mean deviation over all fitted pixels. For our fit, the above parameters are equal to 0.16 and 4.2 respectively. Using the $\rm M_{BH}- \sigma^*$ relation \citep[][]{M.Sigma.Gebhardt} and adopting the constants from \citet[]{M.Sigma.Batiste.2017} derived for the AGN sample, we estimated the black hole mass of J0644$+$1043 as $\rm M_{BH}$=\,4.1$^{+9.39}_{-2.87}$\,$\times$10$^8$ {\msun}. It is rather typical one or little lower than the mean value of $\rm M_{BH}$ for a sample of GRGs obtained by D20: $\rm M_{BH}$=\,1.04\,$\times$\,10$^9$ {\msun}\,or by  \citet[]{5rMpc.GRG.Oei}: $\rm M_{BH}$=\,1.5\,$\times$\,10$^9$ {\msun}.

\subsection{Radio morphology}\label{subsec4.2:result}
The new sensitive uGMRT observation revealed thin, collimated radio jets emanating from the radio core, terminating with detached lobes on either side like a typical FR\,II RG. However, the lobes show a peculiar S-shaped morphology (see Figure\,\ref{fig:uGMRT_map}). The source has a pronounced hotspot at the edge of the eastern lobe (EL). However, the western lobe (WL) has an extended plateau instead of a compact hotspot-like feature. The projected linear size of J0644$+$1043 from the eastern hotspot (EH) to the peak of the western plateau (WP), following the path of the jets, is 711 kpc (12\arcmin, angular size, see the left panel of Figure\,\ref{fig:ugmrt_B3_B4_map}), making this RG a GRG. The core flux density measurements $(S_{\nu,core})$ were obtained using the \texttt{CASA} \texttt{imfit} and the error in the core flux measurements was obtained using $\rm \Delta S_{\nu,core}  = \sqrt{(S_{\nu,core} \times 0.05)^2 + \left(err_{imfit}\right)^2}$. All measurements are listed in Table\,\ref{tab2:flux_comp}. The core remained unresolved at high angular resolution maps we analyzed.

The flux density of the EH is higher than that of the WP. Both lobes have plume-like extensions that start from the hotspot/plateau and extend in a direction nearly 30\arcdeg\,to the jet axis from east to north and west to south. They are completely detached from the jets, and there is no visible connection between the jets and the plumes. The GRG J0644$+$1043 shows a peculiar S-shaped morphology with no evidence of any kind of backflow or recurrent jet activity in the uGMRT maps (see Figure\,\ref{fig:ugmrt_B3_B4_map} and \ref{fig:uGMRT_map}). Hereafter, we refer to J0644$+$1043 as an S-shaped GRG (SGRG). The projected linear sizes of the eastern (from EH to EPE in the left panel of Figure\,\ref{fig:ugmrt_B3_B4_map}) and western (from WP to WPE in the left panel of Figure\,\ref{fig:ugmrt_B3_B4_map}) plumes are 455 (7\arcmin.7) and 200 (3\arcmin.4) kpc, respectively.

\begin{figure*}[ht!]
    \centering
    \includegraphics[scale=0.7]{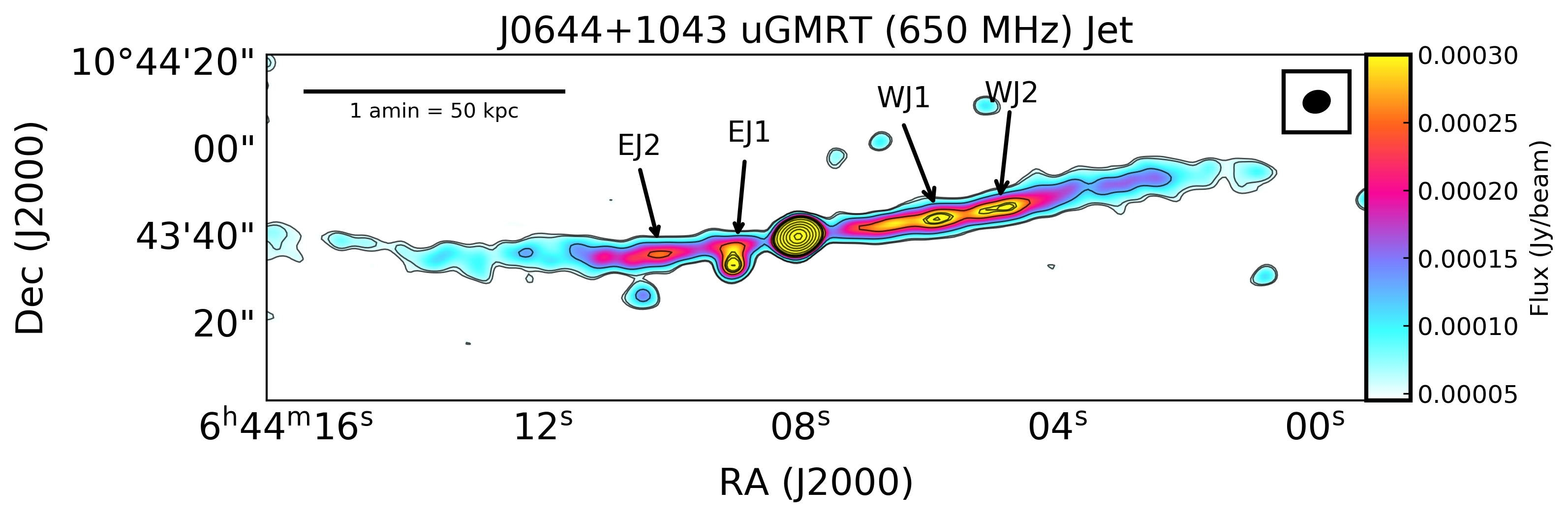}
    \caption{The kpc-scale jet of J0644$+$1043: 650\,MHz uGMRT thin contours superimposed on the color map. The EJ1, EJ2, WJ1 and WJ2 mark the jet knots. The synthesized beam size of  5\farcs6 $\times$ 4\farcs4 is marked by a filled ellipse in the top right corner of the image. The contours have a square-root scaling and the first one is plotted at three sigma level ($\rm \sigma$ = 20 \uJy).}
    \label{fig:jet}
\end{figure*}

We have detected a kpc-scale jet and a counter-jet in the new uGMRT observations. The jets are not straight but slightly curved towards the lobes; the eastern jet's connection to the lobe is not discernible, while the western jet (WJ), which can be seen as separate parts, eventually is feeding to the western lobe. The WJ is bending just before the base of the plume at WL (labelled as JB1 and JB2 in the left panel of Figure\,\ref{fig:ugmrt_B3_B4_map}). The possible knot-like features (marked as EJ1, EJ2, WJ1, and WJ2 in the Figure\,\ref{fig:jet}) are visible on both sides of the radio core. Due to the large beam size and sensitivity limit, these are not visible in the NVSS or RACS. The projected linear sizes of the eastern and western jets are $\sim$ 85 (1\arcmin.45) and 108 (1\arcmin.8)\,kpc, respectively. We have considered the fainter jet (i.e., EJ) as the counter-jet. Both jet and counter-jet meet the jet criteria of \citet[]{Jet.Review.Bridle84}{}. Table\,\ref{tab2:flux_comp} summarizes the integrated flux densities for each component of the SGRG.

Next, we prepared a SI map using tapered uGMRT band 3 and band 4 maps with a common resolution of 25\arcsec. In order to achieve this, we used calibrated output visibilities from all sub-bands obtained from \texttt{SPAM}. These were deconvolved jointly using \texttt{WSClean} with the option of a fixed common beam size for both bands. The other imaging options were the same as those discussed in Section\,\ref{subsec3:radio}. Finally, we obtained two deconvolved images in a similar way as described in Section\,\ref{subsec3:radio} with exactly the same beam size of 25\arcsec $\times$ 25\arcsec. The SI map shows a SI value of $-1\pm0.2$ for the core. In addition, using the flux density measurements of the core (see col. 4 of Table\,\ref{tab2:flux_comp}) the SI between uGMRT band 3 (or band 4) \footnote{The angular resolution of the uGMRT band 3 and 4 are given in Col. 5 of Table \ref{tab1:wsclean_param}} and VLASS (angular resolution of 3\farcs0 $\times$ 2\farcs4) was estimated and the spectrum is also steep with an SI value of $-0.9 \pm 0.1$ (or $-0.8\pm 0.1$). RGs whose core peaks at frequencies below 400\,MHz are called compact steep spectrum (CSS) sources, which typically have sizes between 0.5 and 20 kpc \citep[][]{ODea.Saikia.CSS}. Steep SI of a core of an extended RG usually implies recurrent activity. \citet[][]{Dabhade2023} found that only about 5\% of GRGs appear to show evidence of recurrent jet activity. One of the most studied GRGs with a CSS-type core is 3C\,236, which shows a double-double morphology with a one-sided jet on the VLBI scale \citep[][]{3C236.VLBI}{}{}. Another example is the GRG J1247$+$6723, whose bright core has a peaked spectrum \citep[][]{1245+676.Marecki.PASA,VLBA.Bondi.CSS.2004}. Based on the similarity to other steep-spectrum core RGs mentioned above, the steep spectrum of SGRG's core indicates potential renewal of jet activity within it. The SI of the jets reveal alternating steeper and flatter sections. Moreover, the SI map shows a gradient in SI values, transitioning from steeper spectrum at the plume edges to flatter spectrum in the brighter regions, a pattern similar to what was observed in the GLEAM -- NVSS SI map (see right panel of Figure\,\ref{fig:nvss_SI}). Notably, in the vicinity of the eastern hotspot and the western plateau, the spectra are flatter compared to those at the edges of the plumes. The plume edges exhibit the expected lower SI value gradient, indicating more and more older electron population towards the edges. The SI map shows an unexpectedly flat spectrum region at the southern edges of the western lobe. This may be due to small values of flux density detected in this diffusive region, which results in large spectral index errors. The average SI value is $-1.14 \pm 0.40$, which is smaller than that obtained from the previous SI map. This discrepancy could be due to the fact that the NVSS map does not detect diffuse steep-spectrum plumes and the SI cannot be estimated at these locations. However, we cannot rule out the existence of a spectral curvature in the broad frequency range (0.2--1.4 GHz) where we estimate the two-point SI.

\setlength{\tabcolsep}{10pt}
\begin{table*}[]
    \centering
    \begin{tabular}{cccccccccccc}
    \hline
    Freq &S$\rm_{EL}$ &S$\rm_{WL}$  & S$\rm_{core}$ &S$\rm_{EJ}$ & S$\rm_{WJ}$ & S$\rm_{total}$  &Ref \\
    (MHz) & (mJy) & (mJy) & (mJy) & (mJy) & (mJy) & (mJy) & \\
    (1) & (2) & (3) & (4) & (5) & (6) & (7) & (8) \\
    \toprule
    87.5 & 729.8$\pm$111.9 & 812.9$\pm$109.5 & - &- &- & 1542.7$\pm$154.3 & 1 \\
    118.5 & 556.9 $\pm$72.7 & 606.7$\pm$75.7 & - &- &- & 1163.6$\pm$116.4 & 1 \\
    154.5 & 529.0$\pm$68.9 & 508.7$\pm$63.8 & - &- &- & 1037.7$\pm$103.8 & 1 \\
    200.5 & 484.9$\pm$60.3 & 446.4$\pm$55.7 & - &- &- & 931.3$\pm$93.1 & 1 \\
    400.0 & 342.5$\pm$34.4 & 293.6$\pm$29.5 & 7.4$\pm$0.4 & 5.7$\pm$0.6 &8.5$\pm$0.1 & 657.7$\pm$64.3 & 2 \\
    525.0 & 271.3$\pm$27.2 & 234.0$\pm$23.4 & 5.6$\pm$0.3 & 4.5$\pm$0.4 &9.0$\pm$0.1 & 524.5$\pm$50.9 & 2 \\
    650.0 & 235.4$\pm$23.6 & 206.0$\pm$20.6 & 4.5$\pm$0.2 & 3.7$\pm$0.4 &7.4$\pm$0.2 & 457.0$\pm$44.4 & 2 \\
    887.0 & 212.0$\pm$21.2 & 162.1$\pm$16.2 & 4.5$\pm$0.4 & 3.4$\pm$0.3 &6.7$\pm$0.3 & 378.7$\pm$38.2 & 3 \\
    1400.0 & 158.4$\pm$15.8 & 106.6$\pm$10.7 & 5.6$\pm$0.5 & - & - & 270.6$\pm$27.1 & 4 \\
    3000.0 & - & - & 1.3$\pm$0.1 & - & - & - & 5 \\
    4850.0 & 47.7$\pm$5.0 & 28.4$\pm$3.0 & - & - & - & 76.1$\pm$8.0 & 6 \\\hline
    \end{tabular}
    \caption{Integrated flux densities of different components of the SGRG. In columns 2-7, flux densities of the eastern lobe (S$\rm_{EL}$), western lobe (S$\rm_{WL}$), core (S$\rm_{core}$), eastern jet (S$\rm_{EJ}$), western jet (S$\rm_{WJ}$) and total (S$\rm_{total}$) are given. The total flux calculation includes the flux of the jet wherever visible. References (Ref): (1) GLEAM: \cite{gleam}, (2) this paper, (3) RACS-low: \cite{RACS}, (4) NVSS: \cite{NVSS}, (5) VLASS: \cite{VLASS}, (6) 87GB: \cite{87GB.1991}.}
    \label{tab2:flux_comp}
\end{table*}

\subsection{Magnetic field and spectral age}\label{subsec4.3:JP_model}
The magnetic field strength in the lobes of an RG can be directly estimated when detected simultaneously at radio and X-ray frequencies. The X-ray emission in the lobes is most likely inverse-Compton scattering between the same relativistic electrons that produce the observed radio synchrotron radiation and the cosmic microwave background (CMB) radiation \citep{Harris1979}. In the absence of adequate X-ray observations, using the equipartition assumption to calculate the magnetic field is the most common practice. However, it is also often found that in the lobes of active FR\,II sources, the magnetic field is generally within a factor of a few (2--3) lower than that implied by equipartition \citep[e.g.,][]{Ineson2017,Turner2018}. On the other hand, however, \cite{Konar2019}, who analyzed inactive (outer) lobes of double-double RGs, found that their magnetic field values were close to those obtained by equipartition. We calculated the magnetic field $B\rm_{eq}$ of the lobes of SGRG using the minimum energy arguments, following the prescription of \cite{Lon11}. Our assumptions included the cutoff frequencies of $\nu_{min}$ = 10\,MHz and $\nu_{max}$ = 100\,GHz, the filling factor equal to 1, and the pure electron-positron plasma. The volume of the lobes of the SGRG was approximated by assuming their shape as rotational ellipsoids with semi-diameters of about 170 and 100 arcsec each. This results in magnetic field strengths of $B\rm_{eq}=0.81\pm0.18$ $\mu G$, and $B\rm_{eq}=0.73\pm0.16$ $\mu G$  for the eastern and western lobes, respectively. These values are small, but not extreme, and comparable to the magnetic field strength estimated for other GRGs \citep[e.g.,][]{Mack98}. The magnetic field errors are calculated assuming 50\% errors in the determination of the volume of the lobes. The corresponding equipartition energy density $u\rm_{min}$ is $(0.61\pm0.27)\times10^{-14}$ $\rm J~m^{-3}$ and $(0.50\pm0.22)\times10^{-14}$ $\rm J~m^{-3}$  for the eastern and western lobes, respectively. These values, although small, are within the limits obtained for other similar objects. For example, in a sample of about 50 GRGs, \cite{Saikia.GRS.1999} found that the minimum energy density was in the range from $0.4\times10^{-14}$ $\rm J~m^{-3}$ to $162\times10^{-14}$ $\rm J~m^{-3}$.

The spectra of the SGRG lobes were first fitted with the Jaffe \& Perola \citep[JP,][]{JP73}, Kardashev-Pacholczyk \citep[KP,][]{Kardashev1962,Pacholczyk1970} and continuous injection \citep[CI,][]{Pacholczyk1970} models to calculate the radiative losses. For both lobes, which do not contain any prominent hotspots, the JP model provided the best fit to the data with minimum $\chi^{2}_{red}$ values. Therefore, in the following, we focus only on the results of this model, which describes the time evolution of the emission spectrum of particles with an initial power-law energy distribution characterized by the injection spectral index, $\alpha_{inj}$, and distributed isotropically in the pitch angle with respect to the magnetic field direction. In applying the JP model, the following assumptions were made: (i) the particles follow a constant power-law energy distribution with no re-acceleration of the radiating particles after entering the lobes, (ii) the magnetic field lines are intertwined and the field strength is constant throughout the energy loss process, and (iii) the time of isotropization is short compared to the radiative lifetime for the pitch angles for the injected particles.

\begin{figure*}[ht!]
\includegraphics[scale=0.55]{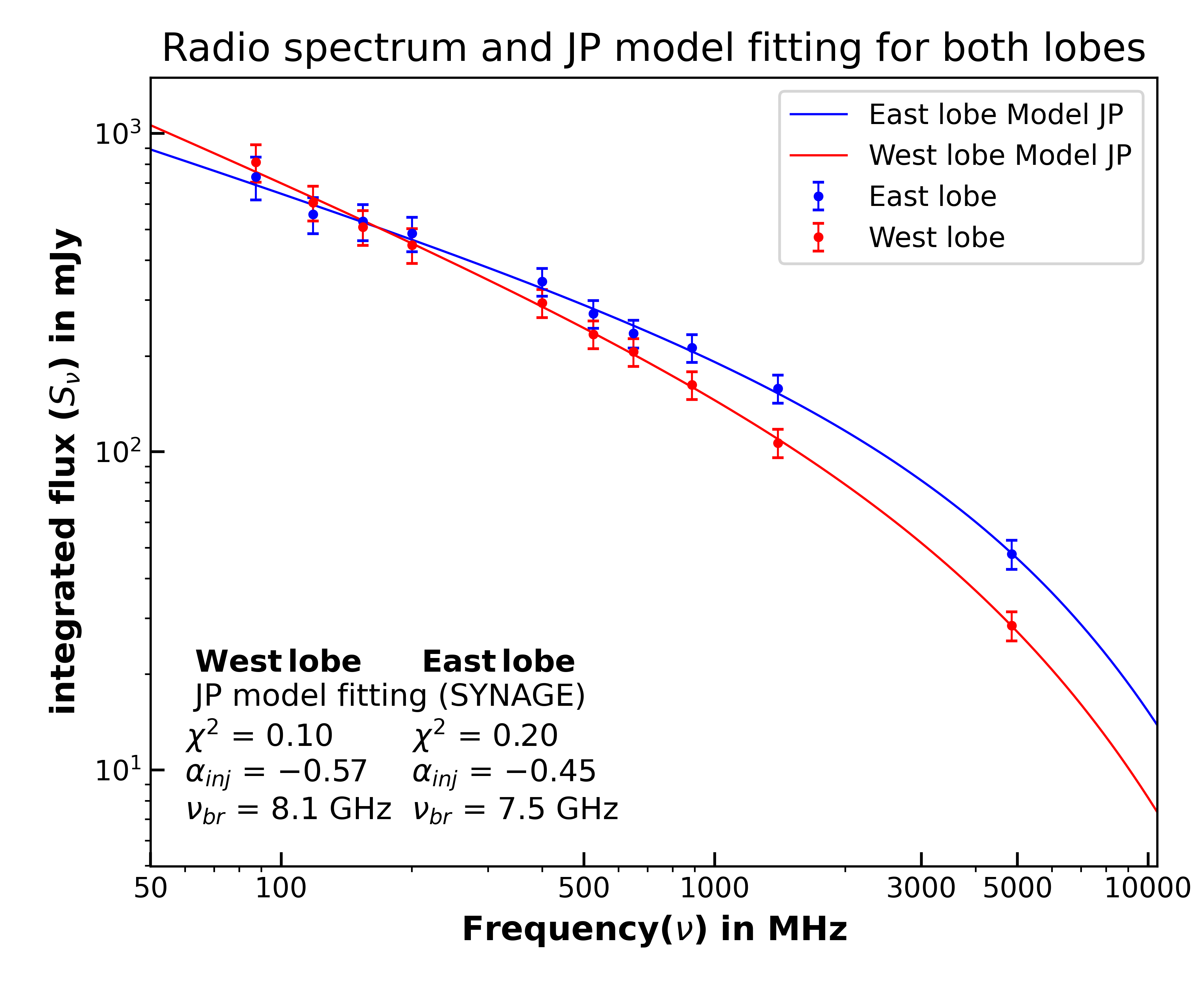}
\includegraphics[scale=0.42]{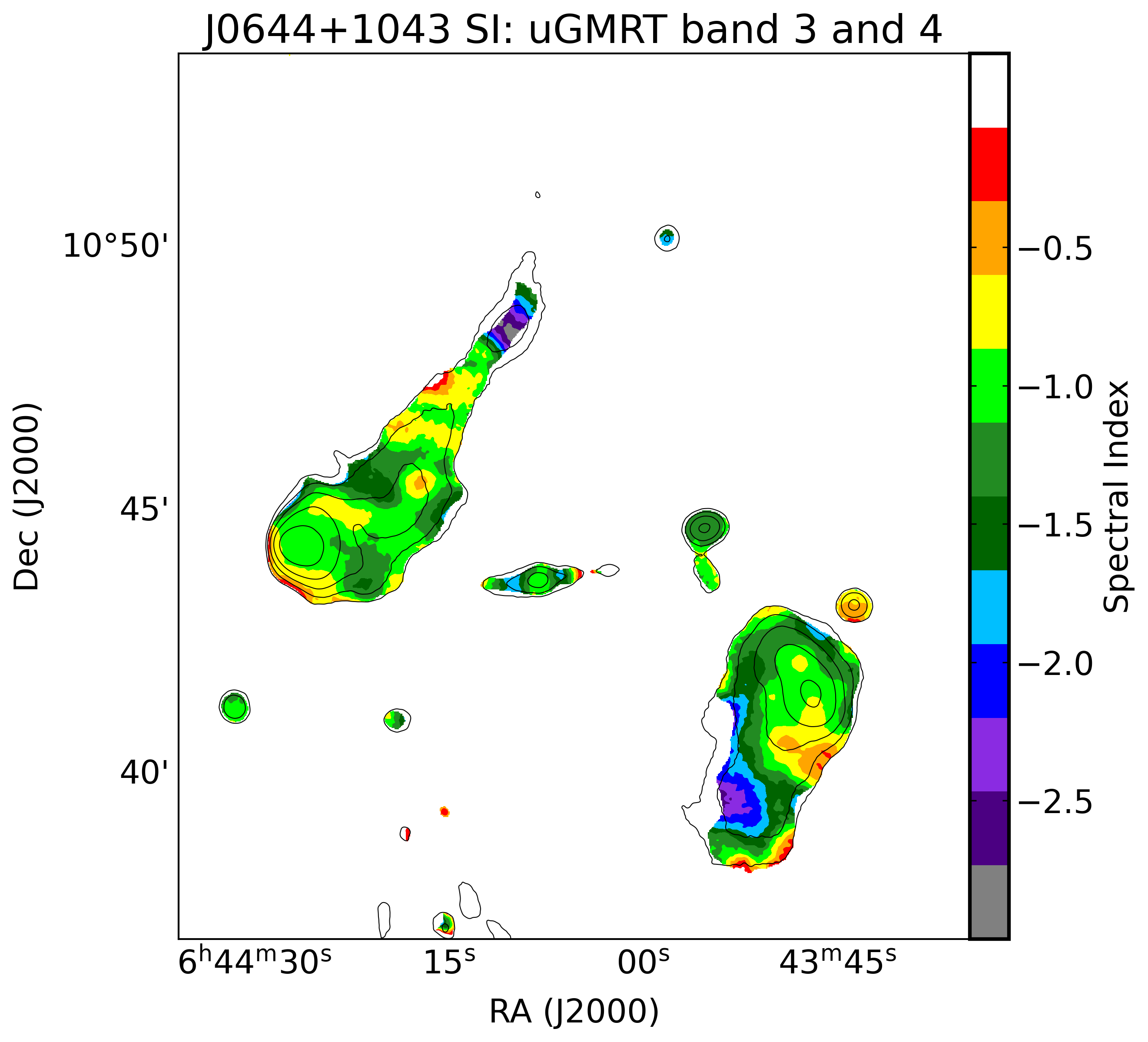}
\caption{The left panel shows the radio spectra of lobes of the target source fitted with the JP model with \texttt{SYNAGE} (for details, see Section\,\ref{subsec4.3:JP_model}). The right panel shows the 25\arcsec\,convolved SI map between uGMRT band 3 and band 4. The uGMRT band 3 contours contours are plotted at $\rm 3\, \sigma \times 2^n$ (n = 0, 1, 2, 3, ...).}
\label{fig:Radio_spectrum_JP_pans}
\end{figure*}

Using the \texttt{SYNAGE} software package \citep{Mur96}, the injection spectral indices of the eastern and western lobes were found to be $\alpha_{inj}=-0.45\pm0.08$ and $\alpha_{inj}=-0.57\pm0.09$, and the break frequencies are of $\nu_{br} = 7.5^{+7.8}_{-2.6}$ GHz and $\nu_{br} = 8.1^{+8.6}_{-2.9}$ GHz, respectively. The best-fit models to the spectra of the lobes using \textbf{\texttt{SYNAGE}} are shown in Figure\,\ref{fig:Radio_spectrum_JP_pans}. Our aging analysis suggests that, for the estimated break frequencies of the eastern and western lobes, the corresponding spectral ages are approximately $38^{+13}_{-12}$ and $35^{+14}_{-12}$ Myr, respectively. The best fitted models to the spectra of the lobes obtained using \texttt{SYNAGE} are shown in Figure\,\ref{fig:Radio_spectrum_JP_pans}. The corresponding $\chi^{2}_{red}$ values of \texttt{SYNAGE} fitting are 0.10  and 0.20 for the western and eastern lobes, respectively. The spectral age is calculated using the following equation:

\begin{equation}
  \tau_{rad}=1590 \frac{B_{eq}^{0.5}}{B_{eq}^2 + B_{CMB}^2} (\nu_{br}(1+z))^{-0.5},
\end{equation}

Where $\tau_{rad}$ is in Myr, the magnetic field is in $\mu$G, and $\nu_{br}$ is in GHz. $B_{CMB}=3.18(1+z)^2$ is the magnetic field strength corresponding to the CMB radiation at our target's redshift $z=0.0488$.

However, it should be noted that the break frequencies associated with the estimated ages are slightly above the frequency range (the highest value is 5 GHz) where the flux densities are measured. This implies that the age of the lobes if we adopt the above value as the break frequency, should be $\lesssim$ 48 Myr. The estimated values of the magnetic field and age of the SGRG are comparable to those of other GRGs analyzed so far. A summary of the literature containing all major spectral ageing studies from the past four decades has recently been provided by \cite{Dabhade2023}. It can be seen (in their Table\,3) that most of the GRGs analyzed have spectral ages $<$ 70 Myr.

\section{Discussion}\label{sec5:discussion}
\subsection{Explanation of the peculiar morphology}
The low-surface brightness wings have a position angle very different from the visible kpc-scale jet direction. This suggests that the central engine has undergone continuous and systematic changes in its spin orientation in the past, manifested by changes in the jet outflow axis. In the introduction, we described some possible mechanisms leading to the disturbed appearance of some RGs. The overall radio morphology of the SGRG resembles an S-shape. Moreover, the well-visible jets are S-wavy themselves, which leads to surmise that the central source/mechanism that produces the jets precess roughly in an anti-clockwise direction with respect to the celestial plane. The jet precession may be due to two circumstances: the presence of another black hole at the same nucleus \citep[e.g.,][]{Begelman.BH.bainary} or a tilted accretion disk \citep[e.g.,][]{Lu.tilted.acc}. From low-resolution optical spectroscopic observations, it is not possible to resolve the possible double-peaked emission lines that may give some hints for binary black holes. However, we cannot completely rule out the possibility of a binary black hole at the center of the galaxy. High-resolution VLBI observations may solve this issue in the future. It is also hard to determine the possibility of a tilted massive accretion disk in the nucleus without thorough X-ray data.

Likewise, the proper motion of the host galaxy with respect to the IGM and/or its orbital motion within a galaxy group can also impact on the apparent structure. However, it is rather unlikely that the orbital motion alone can be responsible for the specific peculiar shape.

We searched for neighboring galaxies in the two-degree (i.e., 7\,Mpc in diameter at z$_{spec}$ = 0.0488) vicinity of the host galaxy and found only 14 galaxies with SDSS photometric redshift resembling the SGRG. We also tried to find any nearby galaxy cluster present near the SGRG host using the catalog of \citet[][]{WHL.2012}, but any catalog of galaxy groups/clusters does not cover this region. However, it should be stressed here that most galaxy-like objects around the SGRG host have no spectroscopic observations and the estimation of photometric redshift is likely to be uncertain due to proximity to the galactic plane. So, the caveat about using photo-z remains.

\begin{figure}[h!]
\includegraphics[scale=0.4]{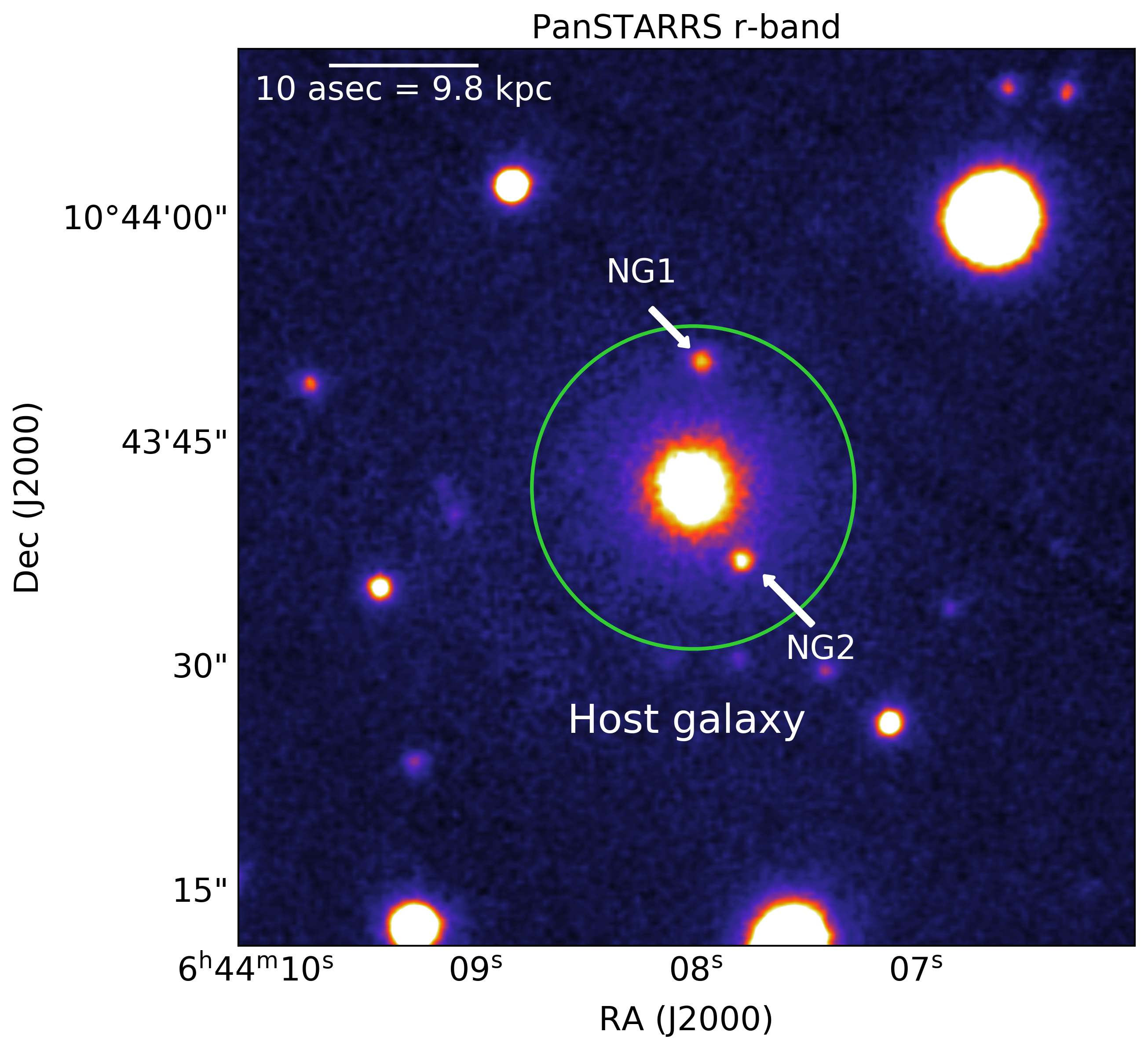}
\caption{The image shows the PanSTARRS r-band image, where the host galaxy is marked with a green-colored circle. The NG1 and NG2 objects are identified as galaxies.}
\label{fig:Pans}
\end{figure}

Interestingly, two objects lie very close to the SGRG host (within an angular distance of $\sim$10$^{\prime\prime}$). They are not listed in the SDSS but appear in the PanSTARRS maps (see Figure\,\ref{fig:Pans}, where they are marked as NG1 and NG2). To determine whether these objects are galaxies or stars, we used a star/galaxy separation method described by \citet{Farrow2014} which based on PanSTARRS magnitude estimates and found that NG1 and NG2 are galaxies. However, they could still be foreground or background galaxies that are not physically related to SGRG. Therefore, the spectroscopic observations of objects nearby SGRG are of high importance.

\subsection{Core prominence}
The prominence of the radio core ($\rm f_c = S_{core}^{5GHz}/S_{total}^{1.4GHz}$) is commonly used as an orientation indicator in studies of beaming in RGs and radio quasars \citep[RQs, e.g.,][]{Orr.Brown82}{}{}. \citet{Jet.Review.Bridle84} found that irrespective of the redshift of an RG, the jet detection rate increases as the relative core prominence increases. Various authors \citep[e.g.,][and reference there in]{Jet3.1.Hardcastle98}{}{} have attempted to show consistency between the distribution of core prominences and the predictions of unified models. \citet[]{Jet.Review.Bridle84}{} found that the core prominences of the nearby 3CR RQs are greater than 0.03. The core prominence value of the SGRG is only 0.0038. This suggests that its nuclear jets should be located close to the plane of the sky. If large-scale jets of the SGRG show a similar orientation in the sky, then we should expect only smaller jet-counterjet asymmetries. Indeed, SGRG shows double-sided kpc-scale jets with similar structure and parameters.

\subsection{Orientation of the kpc-scale jets}
Assuming intrinsically symmetric jets and using the jet-to-counter-jet flux density ratio (J) along with the jet bulk-flow velocity ($\beta_j$), one can determine the jet inclination angle ($\rm \theta$) with respect to the observer’s line of sight. In the past, several attempts have been made to confine the value of the relativistic jet velocity. \citet[]{Wardle.Aaaron.jet.speed} made this assignment for a sample of extended RQs. \citet{Jet7.Mullin09} obtained bulk-flow velocity values between 0.53c and 0.74c using the Bayesian inference method for 92 FR II RGs. However, this speed may be much lower for some individual RGs, e.g., for WAT type RGs, for which \citet{Hardcastle.WAT.2004} obtained the value of 0.3c. Using the recipe provided by \citet[]{Hocuk.Barthel} we calculated $\rm \theta$ using the following equation:

\begin{equation}\label{eq:jet-angle}
\rm \theta = \arccos \left[\cfrac{1}{\beta_j}\,\,\cfrac{s-1}{s+1}\right]
\end{equation}

\begin{figure}[h!]
\includegraphics[scale=0.45]{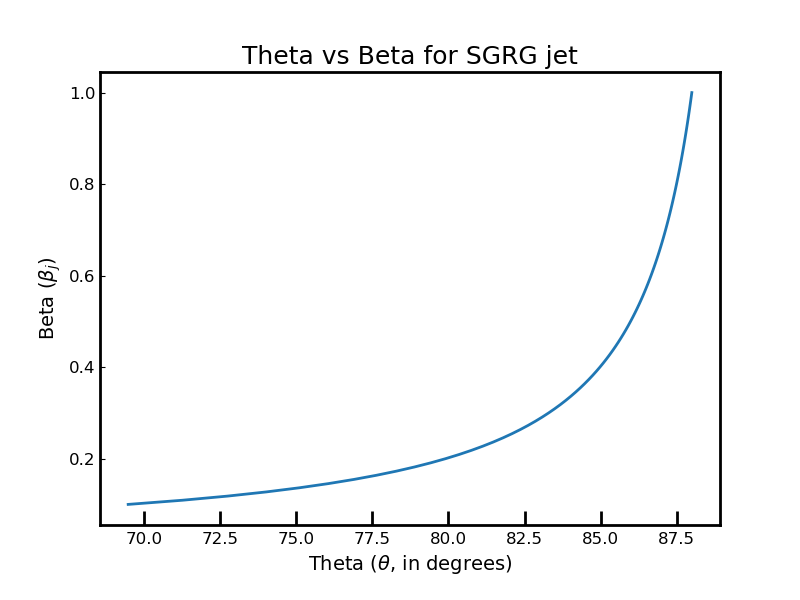}
\caption{The image shows the dependence of the jet bulk flow velocity $\beta_j$ on the jets inclination angle $\theta$ for J0644+1043 (for details see Section\,5.3).}
\label{fig:jet_angle_new}
\end{figure}

Where s = $J^{1/(2-\alpha)}$ and $\alpha$ is the spectral index of the jet. From the 525\,MHz uGMRT map of SGRG, we obtained\footnote{For the calculation of J, we followed the procedure defined by \citet[][]{Jet7.Mullin09}{}{}} $\rm J = 1.2$ and adopting the average spectral index value of $-$0.6, we can depict the $\theta$--$\beta_j$ relation, as illustrated in Figure\,\ref{fig:jet_angle_new}. The resulting jet inclination angle $\theta$ is 88\arcdeg, 87\arcdeg\, and 83\arcdeg\, for $\beta_j$ equal to 0.99c, 0.6c, and 0.3c, respectively. It indicates that the large-scale jets of SGRG are lying close to the plane of the sky.

\begin{table*}[ht!]
    \centering
    \begin{tabular}{ccccc}
    \hline
    Redshift & Spectral & One-sided jet & Two-sided jet &Ref\\
    (z) & class & detection (\%) & detection (\%) &\\
    \toprule
    $<$0.15 & FR II RGs & 62 & 20 &1,2 \\
    0.15$<$z$<$0.3 & FR II RGs  & 76 & 28 &3  \\
    0.3$<$z$<$0.6 & FR II RGs \& RQs & 62 & 7 &4  \\
    0.6$<$z$<$1.0 & FR II RGs \& RQs & 50 & 3 & 5 \\
    \hline
    \end{tabular}
    \caption{Occurance rate of definite or possible kpc-scale long one- and two-sided jet. References (Ref): (1) \citet{Jet1.Black92}, (2) \citet{Jet2.Leahy97}, (3) \citet{Jet3.Hardcastle97}, (4) \citet{Jet4.Gilbert04}, (5) \citet{Jet5.Mullin06}.}
    \label{tab3:jet_occ_final}
\end{table*}

\subsection{Occurrence of large-scale jets}
The jets in SGRG are distinct and well-collimated with a total projected linear size of $\sim$100\,kpc in the sky. The detection rate of such large jets among RGs is not high and is even rarer in GRGs. There are just very few examples of RGs which have detectable total $\sim$ 100\,kpc long collimated jets (e.g., NGC\,315, \citealt[][]{NGC315}; NGC\,6251, \citealt[][]{NGC.6251.Perley.Bridle.84}; HB\,13, \citealt[][]{CGCG.HB13.Jaegers}; CGCG 049$-$033, \citealt[][]{CGCG.Bagchi}; J2233$+$1315, \citealt[][]{Pratik.Barbell}). A statistical study of the occurrence rate of jets was conducted by \citet{Jet.Review.Bridle84}. The authors found that jets are detectable in the range from 65\% to 80\% and from 40\% to 70\% in the case of low-power RGs and powerful RQs, respectively. Based on high-frequency sensitive VLA maps many authors \citep[see e.g.,][and references therein]{Jet6.Mullin08}{}{} have studied the properties of FR II sources including their jets. Pursuant to these analysis we made a compilation of jet occurence for various FR\,II RGs and the results are shown in Table\,\ref{tab3:jet_occ_final}. One can see that while single-sided definite or possible jets occur in over half of the examined objects regardless of their redshift, double-sided definite or possible jets are detectable in about 25\% of low redshift RGs, and in the case of high redshift, the number drops to a few percent only. In this light, J0644$+$1043 with its well-collimated kpc-scale twin jet appears as a rather unusual object.

\subsection{SGRG a low-power FR\,II RG}
The total and upper limits to the core radio power of SGRG are 5.92 $\times$ 10$^{24}$ \whz\, at 1.4\,GHz and 5.86 $\times$ 10$^{21}$ \whz\, at 5\,GHz, respectively. Therefore, SGRG jets are weak-flavored type \citep[][]{Bridle.1984.type.of.jet}, but they terminate with brighter features towards their edges (i.e., the eastern hotspot and the western plateau). The weak jets are generally found in typical weak FR\,I type RGs. However, such weak-flavored jets usually do not manifest any knot-like features, while here they do involve compact structures marked as EJ1, EJ2, WJ1, and WJ2, normally observed in powerful RGs or quasars \citep{Jet.3CR.Bridle94,PKS.knot.Godfrey}. The feature marked as EJ1 in the eastern jet does not extend exactly in the direction of the jet spine-line. It is not clear whether it is related to jet at all. Moreover, the SGRG jets do not flare continuously like jets in typical FR\,I RGs \citep[][]{Jet.review.FRI.Laing14}, instead, they are rather well-collimated after the first stage. Therefore, it is not possible to unambiguously determine which FR type the SGRG is. This gives rise to two possibilities concerning its mixed FR class behavior. The first one states that it is a transitional RG, like NGC\,315 \citep[][]{NGC315} and/or NGC\,6251 \citep[][]{NGC.6251.Perley.Bridle.84}. The other is that the SGRG belongs to a group of FR\,II RGs with weaker jets that need more investigation. In the last decade, our understanding of FR division based on radio luminosity changed drastically with deep radio observations. \citet[]{Mingo.FRI.FRII.2019}{}{} found many low-power FR\,IIs ($\rm P_{Total}^{150\,MHz}  < 10^{26} $ \whz\,) and this indicates that the FR division does not depend solely on the power of the jet. Also, the environment plays a role in forming FR\,I or FR\,II-type jets. This RG could be an example of low-power FR\,IIs that the new generation of surveys is just beginning to reveal. If so, it will be interesting to look into the jet morphology of SGRG with high-frequency deeper subarcsecond resolution observation to check whether its jet is similar to strong-flavored jets (as in powerful FR\,II RGs or RQs) or weak-flavored jets (as in FR\,I RGs).

\section{Conclusion}\label{sec6:conclusion}
Our dedicated uGMRT observations of the RG J0644$+$1043 and the 2-m Rozhen optical telescope detection of its host galaxy revealed previously unknown properties of this source. 
\\
(i) Our spectroscopic observations revealed the redshift of the host galaxy at z=0.0488, which is five times smaller than the photometric redshift previously used in the literature.
\\
(ii) From the low-frequency radio observations we obtained the best possible sensitive map by combining the uGMRT band 3 and band 4 data. The resulting broadband (300--750 MHz) 30 \uJy sensitive map revealed an exceptionally distinctive S-shaped morphology of J0644$+$1043. 
\\
(iii) This unique giant (with a linear size of 0.71~Mpc) low-power ($\sim$6 $\times 10^{24}$ \whz \,at 1.4\,GHz ) radio structure with a synchrotron age of about 50~Myr is thought to be formed due to the systematic precession of the jet of its central supermassive black hole of the mass of $\rm M_{BH}$=\,4.1$^{+9.39}_{-2.87}$\,$\times$10$^8$ {\msun}.
\\
(iv) Our high angular resolution maps revealed a two-sided, wavy, well-collimated, and knotty jet of 100~kpc size. These jets can be considered as ``naked jets'', as no diffuse cocoon was detected around them. Moreover, such large jets are very rare in GRGs.
\\
(v) The core prominence value of 0.0038 and the large-scale jet's sidedness ratio of 1.2 suggest that its nuclear and the kpc-scale jets are located close to the plane of the sky (i.e., 87\arcdeg\, and 83\arcdeg\, for $\beta_j$ equal to 0.6c and 0.3c). The unresolved core has a steep spectrum, indicating a CSS-type nature source.
\\
(vi) The overall radio morphology and energetics of this SGRG and the structure of its jets do not allow a clear determination of the FR-type of this RG.
\\
A better explanation of the nature of this unusual object will require further multi-frequency deep radio observations of its core and jets with subarcsecond resolution.

\begin{acknowledgments}
The authors thank the anonymous reviewer for her/his valuable comments and suggestions. We acknowledge Huib Intema, Andre Offringa, Błażej Nikiel-Wroczyński, Staszek Zola, Alek Kurek, Arti Goyal, Kamil Wolnik, Łukasz Stawarz, Anna D. Kapinska, Frank Schinzel, Rick Perley, Ruta Kale and C. H. Ishwara  Chandra, for their valuable suggestions or help.

We thank the staff of the GMRT that made these observations possible. GMRT is run by the National Centre for Radio Astrophysics of the Tata Institute of Fundamental Research. S.S., A.K., and M.J. were partly supported by the Polish National Science Centre (NCN) grant UMO-2018/29/B/ST9/01793. S.S. also acknowledges the Jagiellonian University grants: 2022-VMM U1U/272/NO/10 and 2022-SDEM U1U/272/NO/15.

The ASKAP radio telescope is part of the Australia Telescope National Facility which is managed by Australia’s national science agency, CSIRO. Operation of ASKAP is funded by the Australian Government with support from the National Collaborative Research Infrastructure Strategy. ASKAP uses the resources of the Pawsey Supercomputing Research Centre. Establishment of ASKAP, the Murchison Radio-astronomy Observatory and the Pawsey Supercomputing Research Centre are initiatives of the Australian Government, with support from the Government of Western Australia and the Science and Industry Endowment Fund. We acknowledge the Wajarri Yamatji people as the traditional owners of the Observatory site. This paper includes archived data obtained through the CSIRO ASKAP Science Data Archive, CASDA (https://data.csiro.au). 

The Pan-STARRS1 Surveys (PS1) and the PS1 public science archive have been made possible through contributions by the Institute for Astronomy, the University of Hawaii, the Pan-STARRS Project Office, the Max-Planck Society and its participating institutes, the Max Planck Institute for Astronomy, Heidelberg and the Max Planck Institute for Extraterrestrial Physics, Garching, The Johns Hopkins University, Durham University, the University of Edinburgh, the Queen's University Belfast, the Harvard-Smithsonian Center for Astrophysics, the Las Cumbres Observatory Global Telescope Network Incorporated, the National Central University of Taiwan, the Space Telescope Science Institute, the National Aeronautics and Space Administration under Grant No. NNX08AR22G issued through the Planetary Science Division of the NASA Science Mission Directorate, the National Science Foundation Grant No. AST-1238877, the University of Maryland, Eotvos Lorand University (ELTE), the Los Alamos National Laboratory, and the Gordon and Betty Moore Foundation.

We acknowledge \textsc{APLpy}\citep[][]{APLpy}, \textsc{astropy} \citep{astropy} and \textsc{matplotlib} \citep{matplotlib} being used in the paper to create all the beautiful plots.
\end{acknowledgments}

\bibliography{ref}{}
\bibliographystyle{aasjournal}

\end{document}